\documentclass[aip,amsmath,amssymb]{revtex4-1}
\usepackage{graphicx}
\usepackage{dcolumn}
\usepackage{bm}
\usepackage{color}
\usepackage{multirow}
\usepackage{colortbl}
\usepackage{subfig}
\usepackage{booktabs}
\usepackage{enumerate}
\usepackage{natbib}
\usepackage{hyperref}
\begin{document}

\title{A Simple Recipe for Estimating Atmospheric Stability Solely Based on Surface-Layer Wind Speed Profile}

\author{Sukanta Basu}%
 \email{s.basu@tudelft.nl}
 \affiliation{Faculty of Civil Engineering and Geosciences, Delft University of Technology, Delft, the Netherlands}

\date{\today}

\begin{abstract}
The wind energy community is gradually recognizing the significance of atmospheric stability in both power production and structural loading. However, estimating stability requires temperature gradient data which are not commonly measured by the wind farm developers or operators. To circumvent this problem, we propose a simple approach \emph{\'{a} la} Swinbank, to estimate stability from only three levels of wind speed measurements. As such, this approach is ideally suited for sodar and lidar--based wind measurements owing to their high vertical resolution in the surface layer.
\end{abstract}


\maketitle

\noindent The Monin-Obukhov similarity theory (MOST) \cite{Monin54}--based surface-layer wind speed profile equation can be written as\cite{Arya01}:
\begin{equation}
U\left(z\right) = \frac{u_*}{k} \left[\ln\left(\frac{z}{z_\circ} \right) - \psi_m\left(\frac{z}{L}\right) + \psi_m\left(\frac{z_\circ}{L}\right)\right],
\label{MOST1}
\end{equation}
where, $\psi_m$ is the so-called stability correction term. This equation has three unknowns: aerodynamic surface roughness ($z_\circ$), Obukhov length ($L$), and friction velocity ($u_*$). Traditionally, either the so-called gradient or the profile method is utilized to estimate these unknowns \cite{Arya01,Emeis13}. Once determined, these micro-meteorological variables can be effectively used in conjunction with Eq.~(\ref{MOST1}) (or one of its generalized versions\cite{Gryning07}) for the vertical extrapolation of wind speeds up to (or higher than) the turbine hub-heights \cite{Motta05,Sathe11,Holtslag17}.  

To the best of our knowledge, with the exception of a multivariate optimization-based approach by Lo\cite{Lo79}, all the other existing gradient or profile methods\cite{Arya01,Nieuwstadt78} require temperature data from two sensor-heights in addition to wind speed measurements. However, in the wind industry, it is not a common practice to measure temperature or temperature gradients. Thus, accurately estimating atmospheric stability, commonly quantified by $L$\cite{Arya01,Motta05}, remains a challenging task in wind resource estimation and other wind energy applications.   

With the advent of remote sensing--based wind measuring instruments (e.g., sodar, lidar), high vertical resolution ($\Delta z$ on the order of a few meters) wind profiles are now abundantly available. In this short communication, we document a simple way to estimate $L$ and other unknowns from Eq.~(\ref{MOST1}) with only three levels of wind speed measurements; absolutely no temperature information is needed. We call this method the hybrid-wind (or, H-W) approach. We explain this approach in a step-by-step manner so that it can be easily implemented by engineers and practitioners outside academia. 

For three wind sensor-heights of $z_1$, $z_2$, and $z_3$, Eq.~(\ref{MOST1}) can be re-written as: 
\begin{subequations}
\begin{equation}
U\left(z_1\right) = \frac{u_*}{k} \left[\ln\left(\frac{z_1}{z_\circ} \right) - \psi_m\left(\frac{z_1}{L}\right) + \psi_m\left(\frac{z_\circ}{L}\right)\right],
\end{equation}
\begin{equation}
U\left(z_2\right) = \frac{u_*}{k} \left[\ln\left(\frac{z_2}{z_\circ} \right) - \psi_m\left(\frac{z_2}{L}\right) + \psi_m\left(\frac{z_\circ}{L}\right)\right],
\end{equation}
\begin{equation}
U\left(z_3\right) = \frac{u_*}{k} \left[\ln\left(\frac{z_3}{z_\circ} \right) - \psi_m\left(\frac{z_3}{L}\right) + \psi_m\left(\frac{z_\circ}{L}\right)\right],
\end{equation}
\end{subequations}
From these equations, the vertical wind speed difference (aka increment) terms can be computed as follows: 
\begin{subequations}
\begin{equation}
\Delta U_{21} = U\left(z_2\right) -  U\left(z_1\right) = \frac{u_*}{k} \left[\ln\left(\frac{z_2}{z_1} \right) - \psi_m\left(\frac{z_2}{L}\right) + \psi_m\left(\frac{z_1}{L}\right)\right],
\label{dU1}
\end{equation}
\begin{equation}
\Delta U_{31} = U\left(z_3\right) -  U\left(z_1\right) = \frac{u_*}{k} \left[\ln\left(\frac{z_3}{z_1} \right) - \psi_m\left(\frac{z_3}{L}\right) + \psi_m\left(\frac{z_1}{L}\right)\right]. 
\label{dU2}
\end{equation}
\end{subequations}
Finally, a ratio of these differences can be written as: 
\begin{equation}
R = \frac{\Delta U_{31}}{\Delta U_{21}} = \frac{\ln\left(\frac{z_3}{z_1} \right) - \psi_m\left(\frac{z_3}{L}\right) + \psi_m\left(\frac{z_1}{L}\right)}{\ln\left(\frac{z_2}{z_1} \right) - \psi_m\left(\frac{z_2}{L}\right) + \psi_m\left(\frac{z_1}{L}\right)}
\label{Ratio}
\end{equation}
It is needless to point out that $R$ is a (nonlinear) function of only $L$. Thus, if $R$ varies in a monotonic manner with respect to $L$, it will be straightforward to estimate $L$ from measured $R$ values via Eq.~(\ref{Ratio}). 
    
The behavior of $R$ depends entirely on the stability correction term ($\psi_m$). For neutral condition (i.e., $\frac{z}{L} = 0$), $\psi_m$ equals to zero. In this case, $R$ is simply a function of three sensor heights: 
\begin{equation}
R_N = \frac{\ln\left(\frac{z_3}{z_1} \right)}{\ln\left(\frac{z_2}{z_1} \right)}
\label{RatioN}
\end{equation} 
Assuming $z_3 > z_2 > z_1$, it is trivial to show that $R_N > 1$. By using well-accepted $\psi_m$ functions, it is also not difficult to show that $R$ is larger (smaller) than $R_N$ for stable (unstable) conditions. 

\begin{figure}[ht]
\centerline{\includegraphics[height=2.7in]{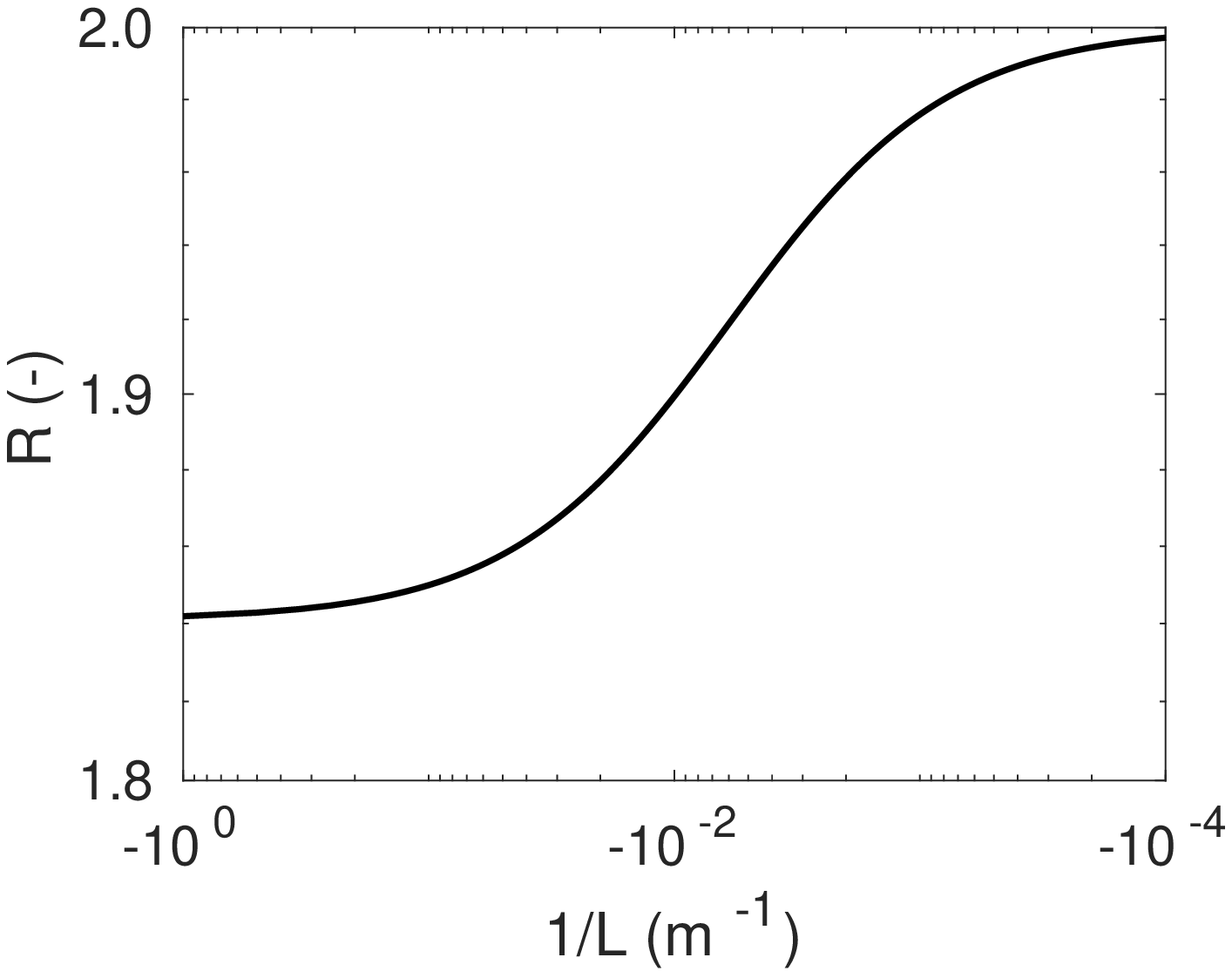}
\includegraphics[height=2.7in]{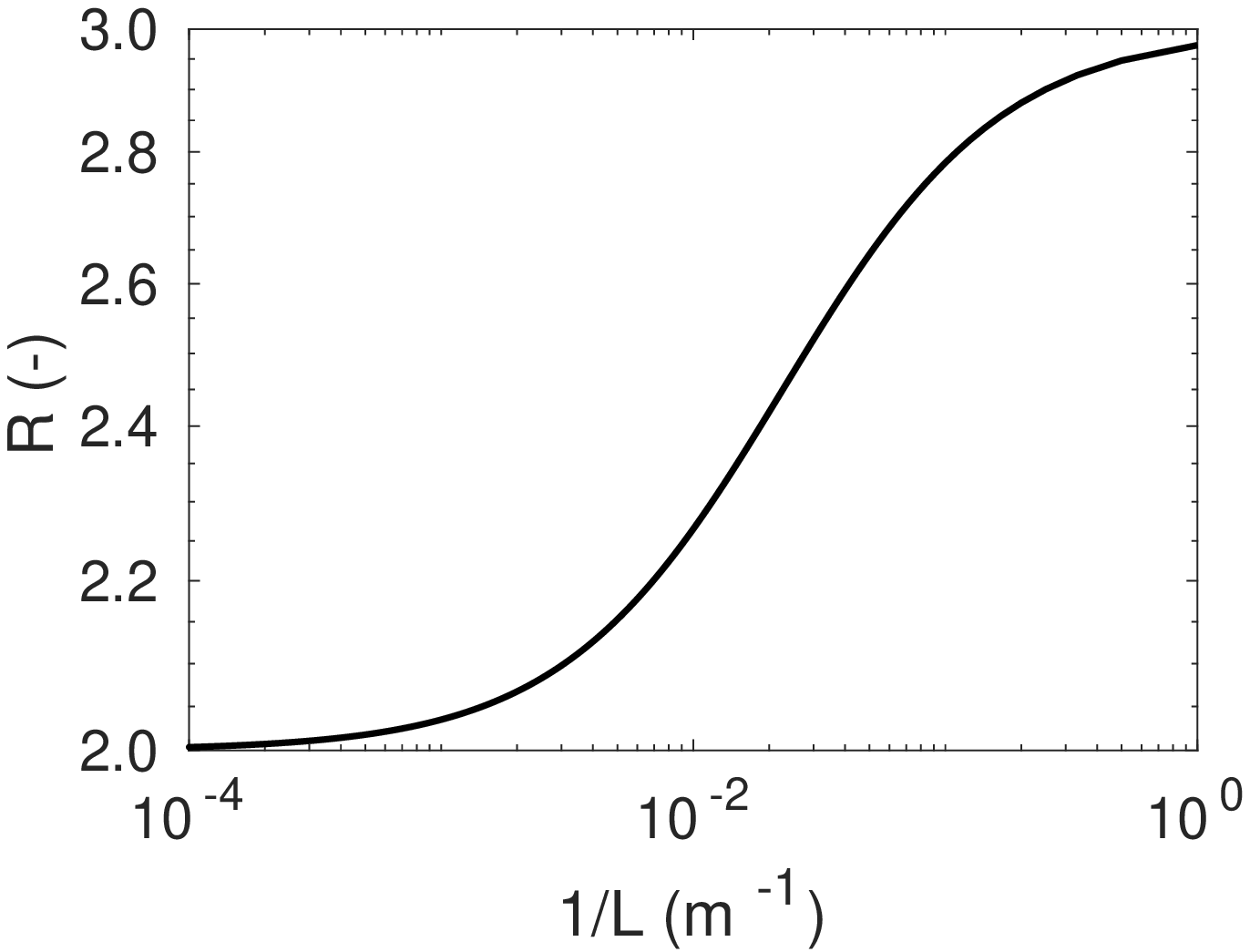}}
\caption{Variation of $R$ with respect to inverse Obukhov length ($1/L$). The left and right panels represent unstable and stable conditions, respectively. The $\psi_m$ formulations by Businger and Dyer [i.e., Eqs.~(\ref{psim1}) and (\ref{psim2})] are utilized here. In these illustrations, $z_1, z_2, z_3$ are assumed to be equal to 5 m, 10 m, and 20 m, respectively. For near-neutral condition ($1/L \rightarrow 0$), $R$ asymptotically approaches 2. \label{F1}}
\end{figure}

The most popular $\psi_m$ functions, attributed to Businger and Dyer \cite{Dyer70,Businger71,Dyer74}, are formulated as:  
\begin{subequations}
\begin{equation}
\psi_m = 2\ln\left(\frac{1+x}{2} \right) + \ln\left( \frac{1+x^2}{2} \right) - 2\tan^{-1}x + \frac{\pi}{2}; \hspace{0.2in} \mbox{for } \frac{z}{L} \le 0
\label{psim1}
\end{equation}
\begin{equation}
\psi_m = -\frac{5z}{L}; \hspace{0.2in} \mbox{for } \frac{z}{L} \ge 0
\label{psim2}
\end{equation}
\end{subequations}
where, $x = \left( 1 - \frac{16z}{L}\right)^{1/4}$. The variation of $R$ with respect to $1/L$ is portrayed in Fig.~\ref{F1}. As an illustration, the sensor heights are assumed to be at 5 m, 10 m, and 20 m, respectively. For these specific height values, $R_N = 2$. Clearly, for unstable conditions (left panel of Fig.~\ref{F1}), $R$ monotonically decreases with increasing instability. In contrast, for stable conditions (right panel of Fig.~\ref{F1}), $R$ shows a monotonically increasing trend with increase in stability.     

Given the monotonic behavior of $R$ with respect to $1/L$, as depicted in Fig.~\ref{F1}, one can easily estimate $L$ given any measured value of $R$. In this regard, any suitable root finding algorithm (e.g., the Levenberg-Marquardt approach) can be utilized in conjunction with Eq.~(\ref{Ratio}). In a nutshell, the proposed H-W approach for stability estimation can be summarized in three steps: 
\begin{enumerate}[1.]
\item Compute $R$ based on measured $U(z_1)$, $U(z_2)$, and $U(z_3)$. 
\item Given $z_1$, $z_2$, and $z_3$, calculate the value of $R_N$ using Eq.~(\ref{RatioN}).
\item If $R < R_N$, then use Eq.~(\ref{Ratio}) in conjunction with Eq.~(\ref{psim1}) to estimate $L$. Conversely, if $R > R_N$, then make use of Eq.~(\ref{psim2}) instead of Eq.~(\ref{psim1}). 
\end{enumerate}

Once Obukhov length ($L$) is estimated, one can estimate the friction velocity ($u_*$) from Eqs.~(\ref{dU1}) and (\ref{dU2}). Since there are two equations and only one unknown, the conventional linear regression approach with ordinary least squares could be employed. Now, $L$ is defined as: 
\begin{equation}
L =  -\frac{\Theta_\circ u_*^3}{k g (\overline{w\theta})}
\label{OB}
\end{equation}
where, $\overline{w\theta}$ is the surface sensible heat flux. The von K\'{a}rm\'{a}n constant is denoted by $k \left(= 0.4\right)$; $g$ is the well-known gravitational constant and $\Theta_\circ$ is a reference temperature (often assumed to be equal to 300 K). After estimating $L$ and $u_*$, Eq.~(\ref{OB}) can be inverted to estimate $\overline{w\theta}$. In other words, both the turbulent momentum and sensible heat fluxes can be (indirectly) estimated using the H-W approach. 

We would like to note that the H-W approach can be further simplified if one has access to reliable aerodynamic roughness value ($z_\circ$). Under this circumstance, only two levels of wind speed data will be required as by definition $U(z_\circ)=0$. Thence, Eqs.~(\ref{Ratio}) and (\ref{RatioN}) become, respectively:  
\begin{equation}
R_* = \frac{U_3}{U_2} = \frac{\ln\left(\frac{z_3}{z_\circ} \right) - \psi_m\left(\frac{z_3}{L}\right) + \psi_m\left(\frac{z_\circ}{L}\right)}{\ln\left(\frac{z_2}{z_\circ} \right) - \psi_m\left(\frac{z_2}{L}\right) + \psi_m\left(\frac{z_\circ}{L}\right)}
\label{RatioBulk}
\end{equation}

\begin{equation}
R_{*N} = \frac{\ln\left(\frac{z_3}{z_\circ} \right)}{\ln\left(\frac{z_2}{z_\circ} \right)}
\label{RatioNBulk}
\end{equation} 
The rest of the procedure remains the same as elaborated before.  

Before demonstrating the capabilities of the H-W approach, some potential pitfalls regarding its usage should be mentioned. First, MOST \cite{Monin54} is strictly valid in a horizontally homogeneous surface layer (where the Coriolis effects can be neglected). In the surface layer (aka constant flux layer), the turbulent fluxes are assumed to be invariant with height. Thus, all the sensor heights (i.e., $z_1$, $z_2$, $z_3$) should be within the surface layer to avoid violation of MOST. For strongly stratified conditions, the surface layer could be only a few meters deep; the H-W approach should be avoided under that scenario. Second, the H-W approach implicitly assumes that wind speed values increase with height. If such condition is not met, it should not be used. 

\begin{figure}[ht]
\centerline{\includegraphics[height=2.7in]{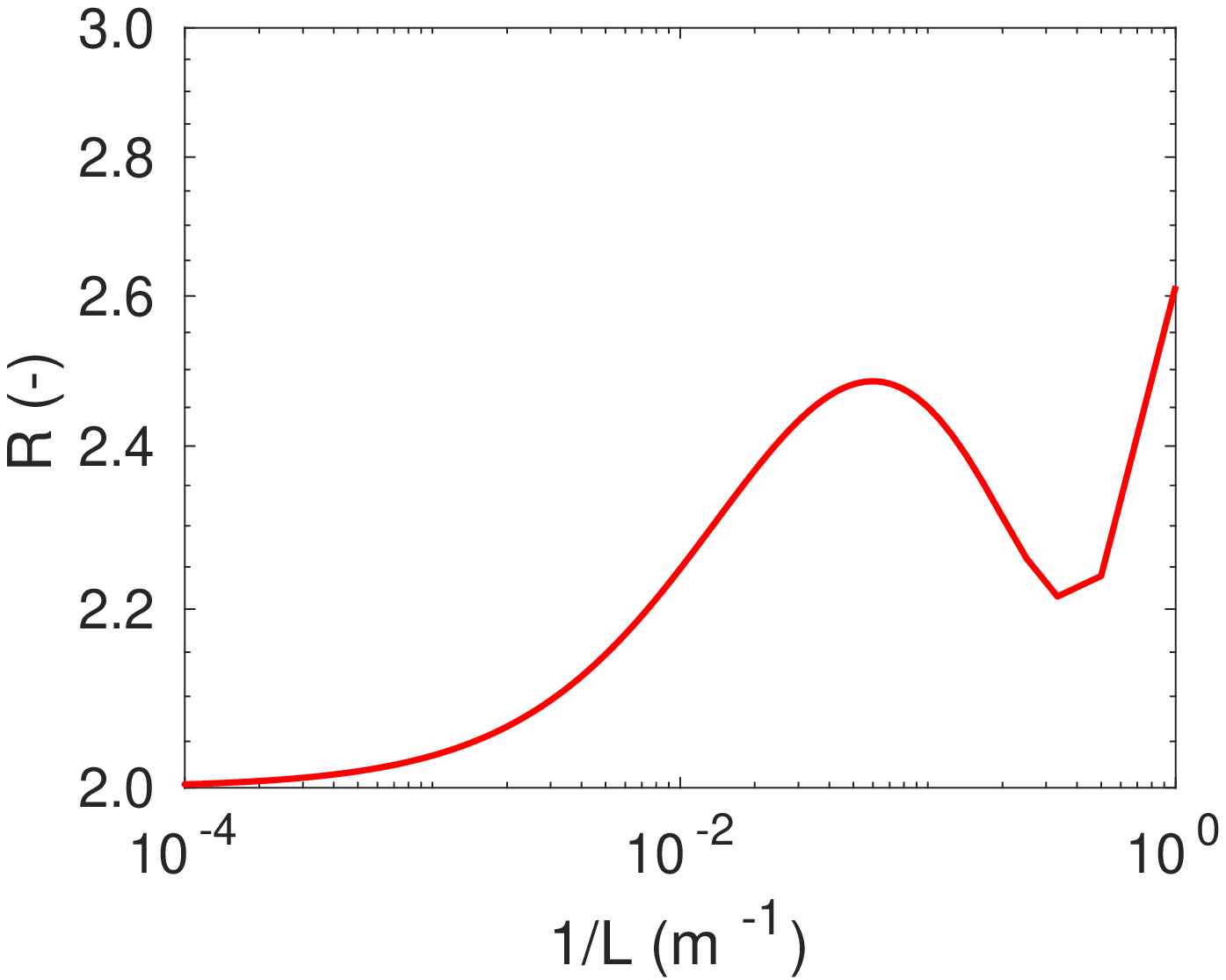}
\includegraphics[height=2.7in]{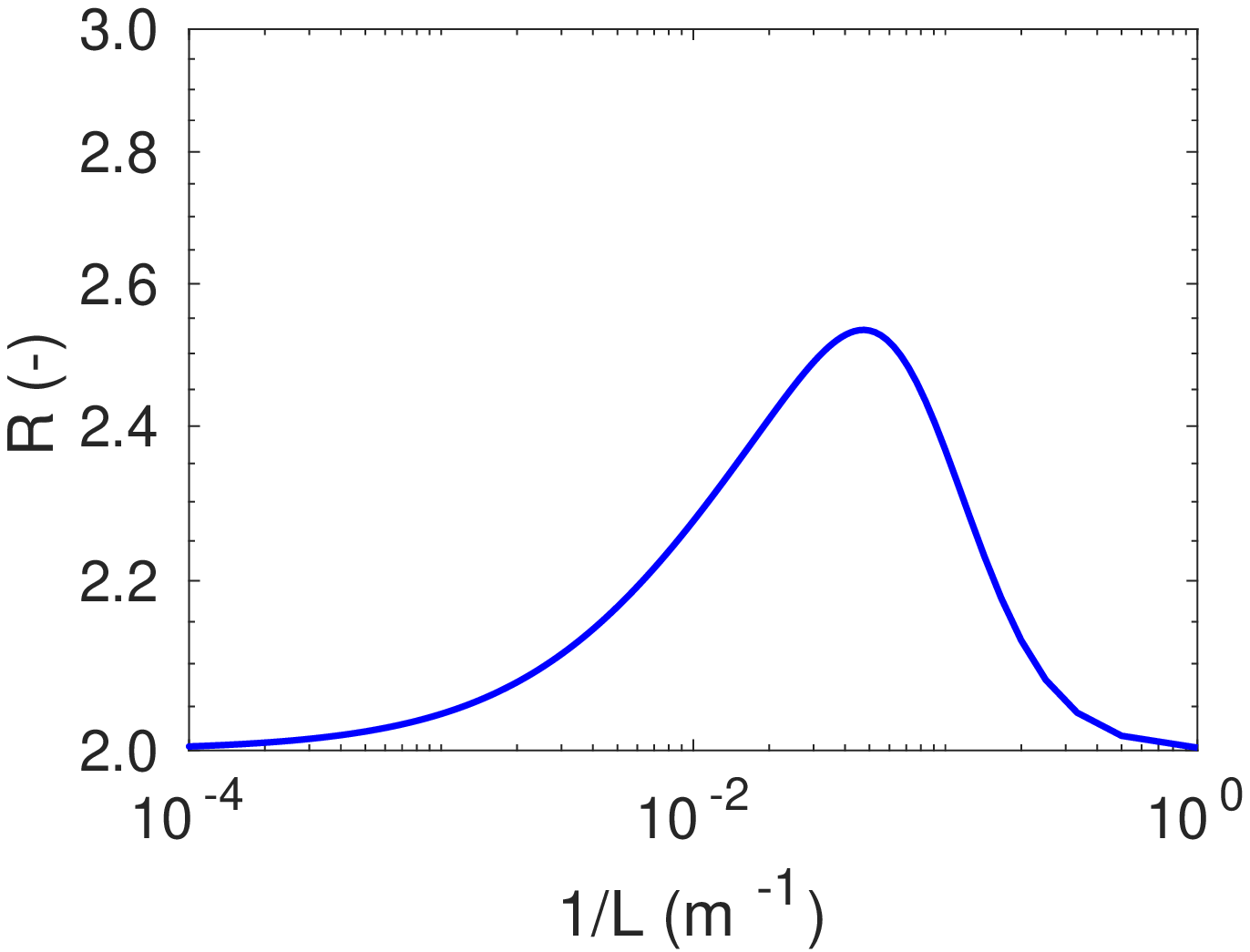}}
\caption{Variation of $R$ with respect to inverse Obukhov length ($1/L$). The left and right panels represent $\psi_m$ formulations by Beljaars and Holtslag \cite{Beljaars91} [Eq.~(\ref{psiBH})] and Cheng and Brutsaert \cite{Cheng05} [Eq.~(\ref{psiCB})], respectively. In these illustrations, $z_1, z_2, z_3$ are assumed to be equal to 5 m, 10 m, and 20 m, respectively. \label{F2}}
\end{figure}

Last, we strongly recommend the usage of stability correction functions ($\psi_m$) proposed by Businger and Dyer \cite{Dyer70,Businger71,Dyer74} while employing the H-W approach. Over the years, several other $\psi_m$ formulations have been proposed in the literature. Most of these formulations disagree among themselves for moderately and strongly stable conditions; for other stability conditions, the consensus is generally very good. For stable condition, the formulation by Beljaars and Holtslag \cite{Beljaars91} reads as: 
\begin{equation}
\psi_m = -a\frac{z}{L} -b\left(\frac{z}{L} - \frac{c}{d} \right)\exp\left(-d\frac{z}{L} \right) -\frac{bc}{d}; \hspace{0.2in} \mbox{for } \frac{z}{L} \ge 0
\label{psiBH}
\end{equation}
where, $a = 1$, $b = \frac{2}{3}$, $c = 5$, and $d = 0.35$. For the same stability regime, Cheng and Brutsaert\cite{Cheng05} proposed: 
\begin{equation}
\psi_m = -a \ln\left[ \frac{z}{L} + \left(1 + \left(\frac{z}{L} \right)^b \right)^{\frac{1}{b}} \right]; \hspace{0.2in} \mbox{for } \frac{z}{L} \ge 0
\label{psiCB}
\end{equation}
where, $a = 6.1$, $b = 2.5$. The $R$-vs-$1/L$ plots based on these formulations are shown in Fig.~\ref{F2}. In contrast to the Businger-Dyer formulation-based plot (right panel of Fig.~\ref{F1}), these plots do not show monotonic behavior (for $L < 20$ m or so). In other words, multiple roots are possible for a given $R$ value. As a result, $L$ cannot be estimated unequivocally.  

In order to validate the H-W approach, we analyze multi-year (2001-2016) meteorological observations measured on the 200~m tall Cabauw tower in the Netherlands \cite{vanUlden96}. Even though the landscape at Cabauw is quite flat and open (grassland), the existence of wind breaks and scattered villages cause significant disturbances in the near-surface region \cite{Verkaik07}. Thus, the Cabauw site is not an ideal location to test any  approach which relies on MOST. Nonetheless, owing to its high-quality, it has been utilized in numerous MOST-related publications\cite{Nieuwstadt78,Holtslag84,Optis14}; we also follow suit. 

Since, at Cabauw, the daytime mixed layer depth is on average significantly lower than 150~m during the winter months (November--January), we exclude data from these months from our analyses. In addition, we discard data from morning (3--9 UTC) and evening (15--21 UTC) transitional periods because surface layer and upper part of the boundary layer often correspond to different stability regimes. We utilize wind speed data from the lowest 3 sensor levels ($z$ = 10, 20, and 40 m\footnote{During nighttime, the surface layer is usually shallower than $z = 40$ m. As such, MOST is not valid at this height for moderately and strongly stratified conditions (categories $g$ and $h$, respectively). Thence, our findings for these categories should be regarded with caution.}) of the Cabauw tower, to estimate $R$. To reduce random errors, we make use of vertical profiles averaged over 30 minutes duration. We do not consider cases where wind does not increase monotonically with height; furthermore, we also exclude the weak wind cases (i.e., $U(z) < 1$ m s$^{-1}$). After imposing all these constraints, we are left with 84,890 profiles for analyses. Based on Table~\ref{T1}, we classify these profiles into several stability categories. These categories were originally proposed by Holtslag\cite{Holtslag84} based on Obukhov length. With the aid of Fig.~\ref{F1}, we convert them to corresponding ranges of $R$. 

\begin{table}[ht]
\centering
\caption{Stability Classification}
\label{T1}
\begin{tabular}{ccc}
\hline
\textbf{Category} & \textbf{$L$ (m)} & \textbf{$R$ (-)} \\
\hline
$a$ & $-40 \le L < -12$     &  $1.8464 < R \le 1.8578$ \\
$b$ & $-200 \le L < -40$    &  $1.8578 < R \le 1.8994$ \\
$c$ & $-1000 \le L < -200$  &  $1.8994 < R \le 1.9583$ \\
$d$ & $|L| > 1000$          &  $1.9583 < R < 2.0673$ \\
$e$ & $200 < L \le 1000$    &  $2.0673 \le R < 2.2651$ \\
$f$ & $100 < L \le 200$     &  $2.2651 \le R < 2.4191$ \\
$g$ & $40 < L \le 100$      &  $2.4191 \le R < 2.6433$ \\
$h$ & $10 < L \le 40$       &  $2.6433 \le R < 2.8782$\\    
\hline
\end{tabular}
\end{table}

\begin{figure}[ht]
\centering
\includegraphics[width=2.8in]{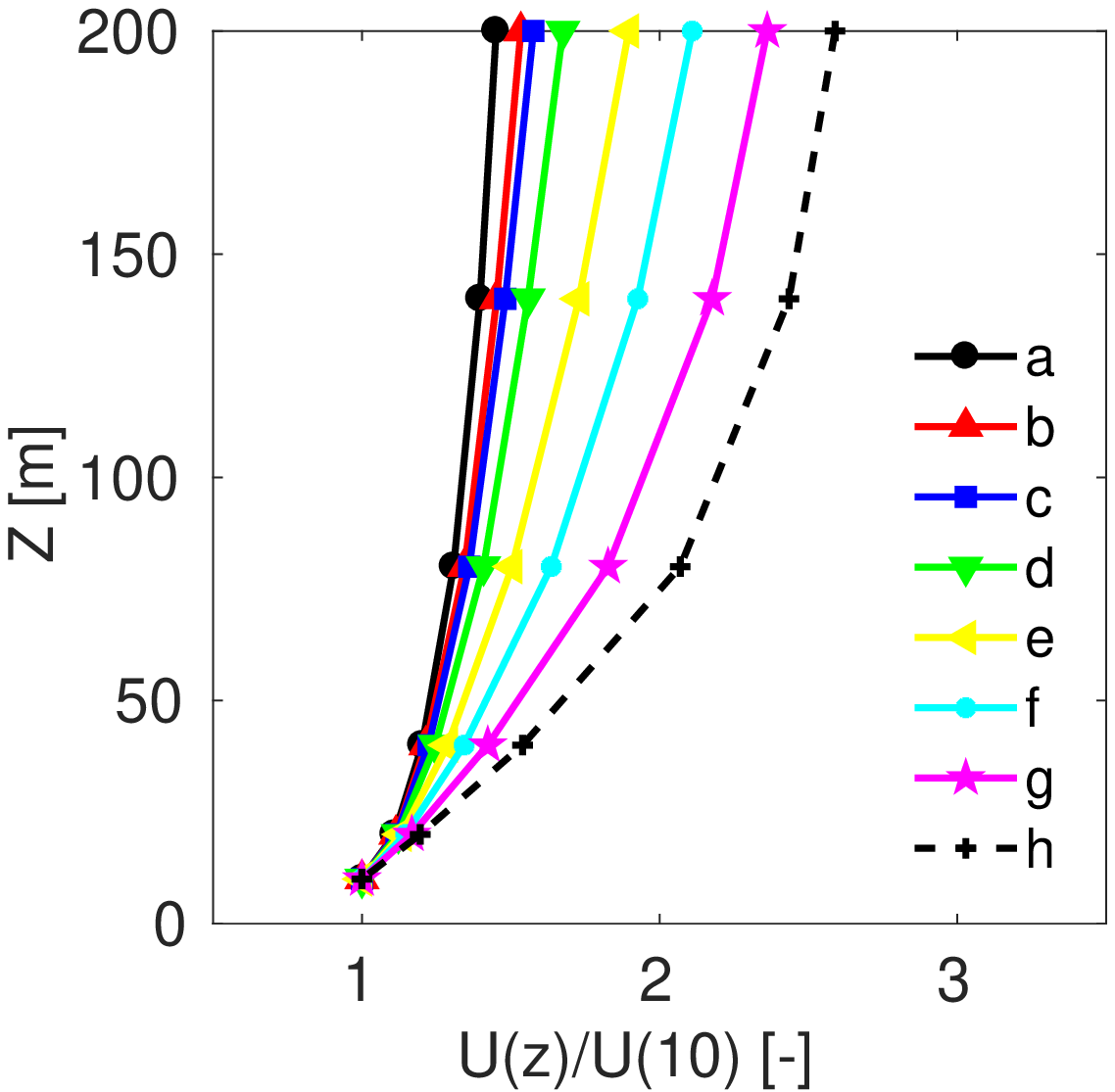}
\hspace{0.3in}
\includegraphics[width=2.8in]{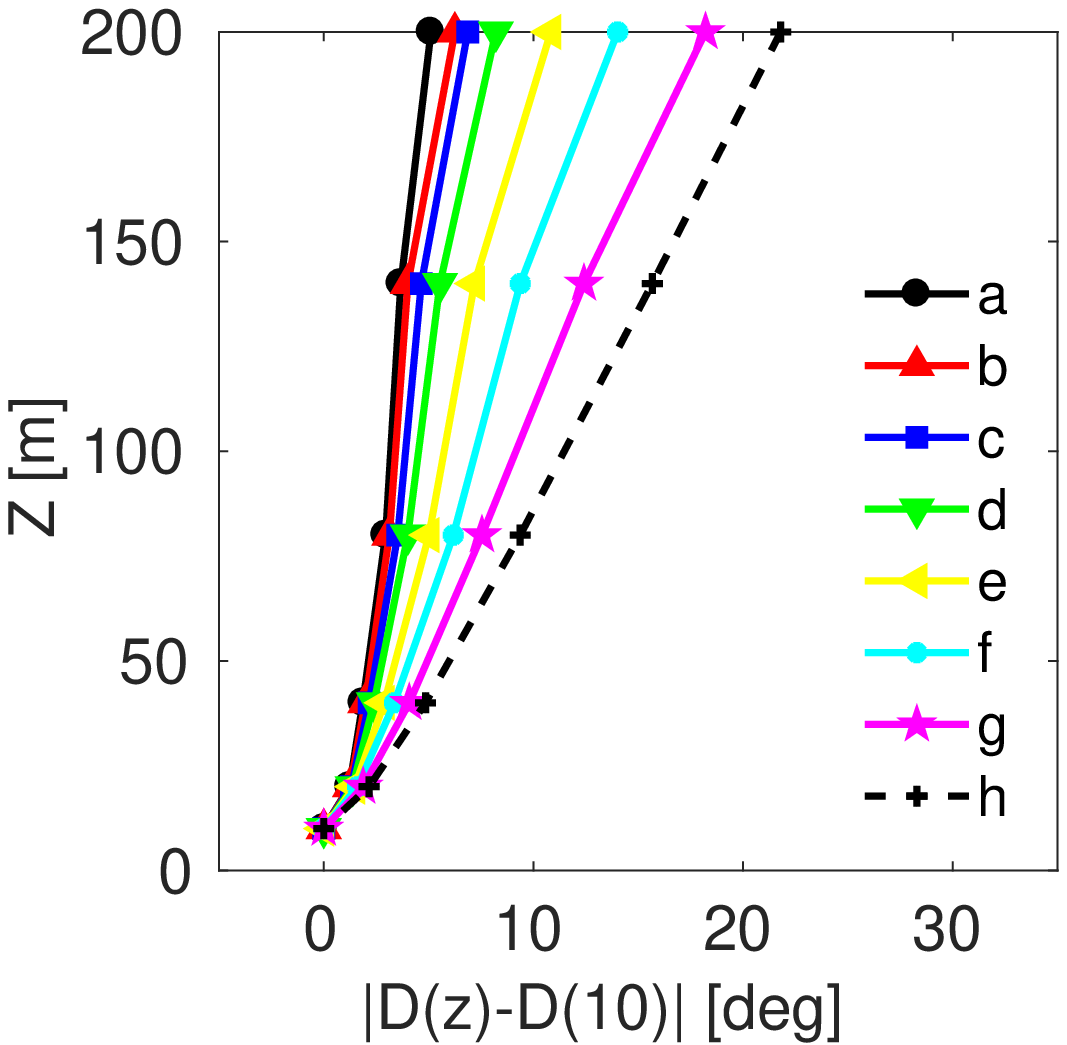}\\
\vspace{0.3in}
\includegraphics[width=2.8in]{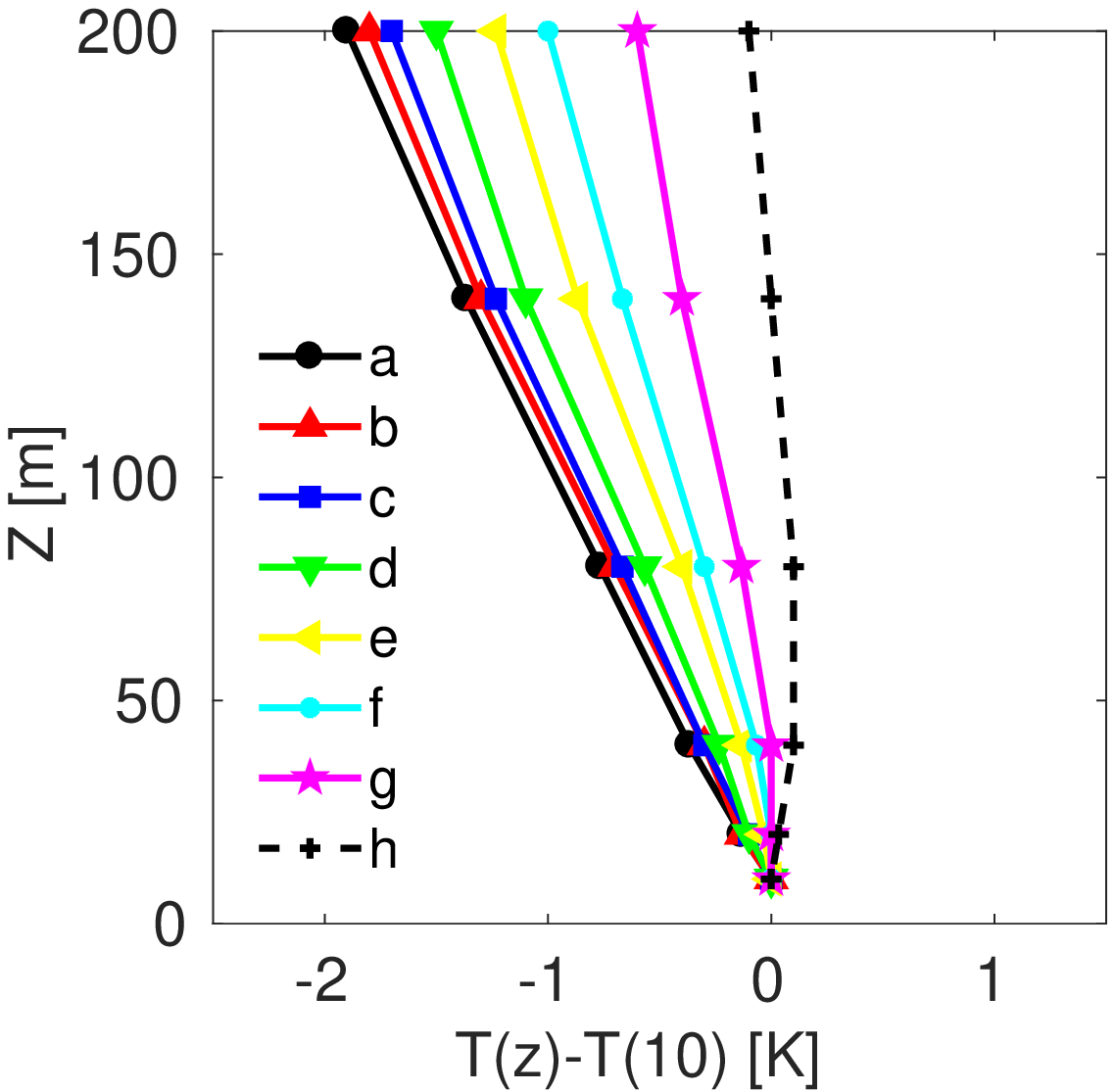}
\hspace{0.3in}
\includegraphics[width=2.8in]{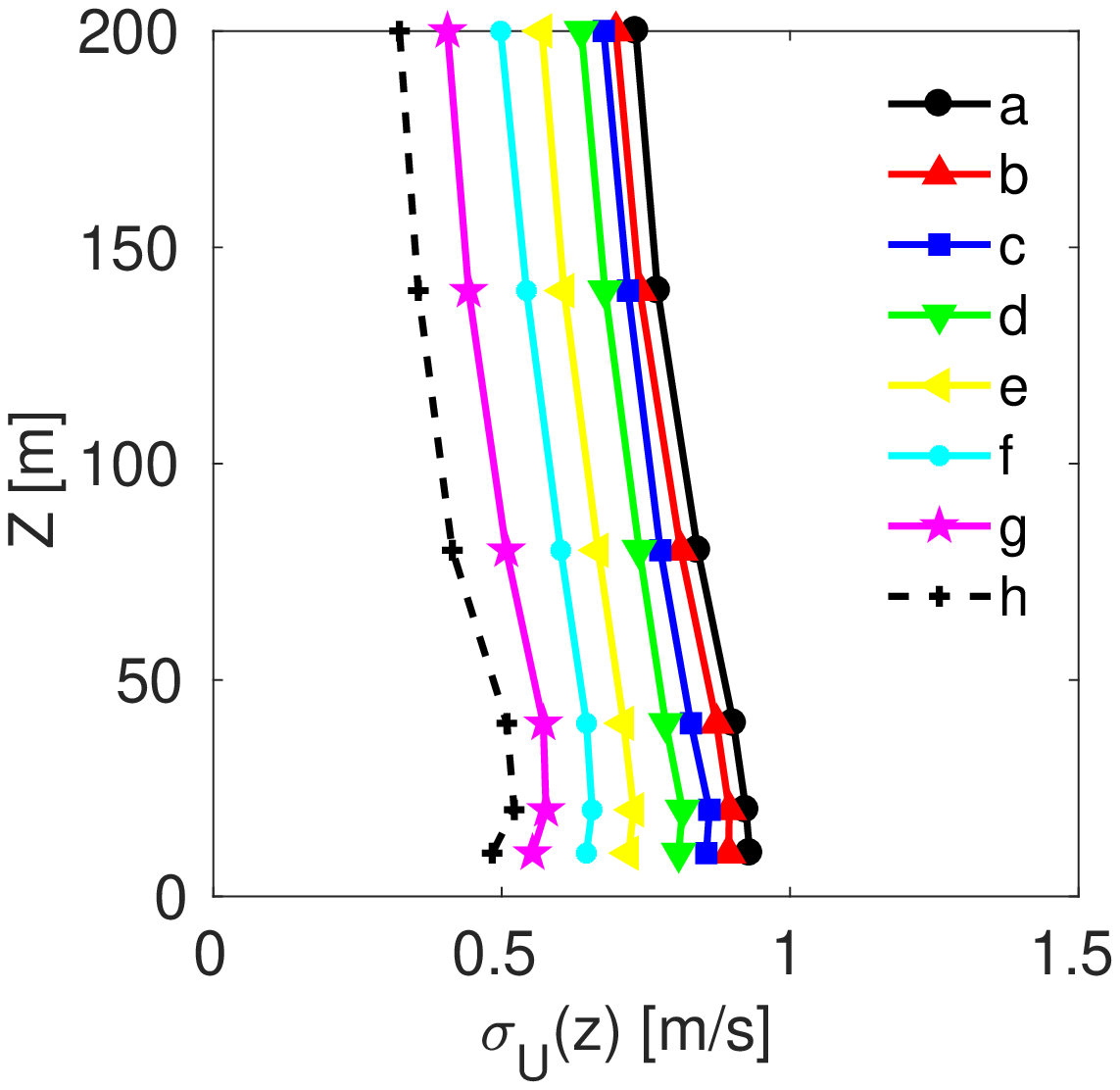}
\caption{Classification of the Cabauw tower-based meteorological observations. Only the wind speed data from the lowest 3 sensor levels are used to estimate Obukhov length. For each stability class, the median profiles of normalized wind speed (top-left panel), wind directional shear (top-right panel), relative temperature (bottom-left panel), and standard-deviation of horizontal wind speed (bottom-right panel) are shown. \label{F3}}
\end{figure}

The median profiles corresponding to each stability class for four key meteorological variables are depicted in Fig.~\ref{F3}. These profiles follow clear trends in agreement with the boundary-layer meteorology literature. For example, both wind speed shear and directional shear values increase with increasing stability\cite{Holtslag84,vanUlden85}. For stability class $a$, the temperature profile closely follows the dry adiabatic lapse rate (-$9.8\times10^{-3}$ K~m $^{-1}$); with increasing stability, the temperature gradients increase monotonically\cite{Arya01}. The decrease in the standard-deviations of horizontal wind speed with increasing stability is also physically realistic \cite{Emeis13}. Thus, overall, the proposed H-W approach has the discriminatory power to classify meteorological profiles into appropriate stability classes. In the future, measured turbulent fluxes from different sites will be utilized to provide more direct validation of the H-W approach.

Before concluding this article, we would like to call the readers'  attention to an old paper by Swinbank\cite{Swinbank64}. In this work, an `exponential wind profile' with a radically different physical basis than the MOST-based Eq.~(\ref{MOST1}) was proposed. In the decade following its publication (as soon as the early 1970s), Swinbank's equation was completely overshadowed by Eq.~(\ref{MOST1}) and forgotten by the boundary-layer meteorology community at large. We stumbled upon this paper serendipitously and were surprised to find out that Swinbank\cite{Swinbank64} outlined an approach virtually identical to the proposed H-W approach, albeit utilizing his `exponential wind profile'. If we are not mistaken, no one else followed-up on his approach or used it in conjunction with Eq.~(\ref{MOST1}). The author of the current article has independently `re-invented the wheel' more than half-a-century later.


\end{document}